\documentclass[acmtog, nonacm]{acmart}
\acmSubmissionID{1074}

\usepackage{booktabs} 

\newcommand{\xhdr}[1]{{\noindent\bfseries #1}.}

\usepackage[ruled]{algorithm2e} 

\SetAlFnt{\small}
\SetAlCapFnt{\small}
\SetAlCapNameFnt{\small}
\SetAlCapHSkip{0pt}

\makeatletter
\renewcommand{\@authorsaddresses}{\par\noindent
Authors' addresses: Jingwen Ye, jingwenye@tencent.com, Tencent AIPD, Shenzhen, China; Yuze He, hyz22@mails.tsinghua.edu.cn, Tsinghua University, Beijing, China; Yanning Zhou, amandayzhou@tencent.com, Tencent AIPD, Shenzhen, China; Yong-Jin Liu, liuyongjin@tsinghua.edu.cn, Tsinghua University, Beijing, China; Xiao Han, elehanx@gmail.com, Tencent AIPD, Shenzhen, China
}
\makeatother

\begin{document}
\title{PrimitiveAnything: Human-Crafted 3D Primitive Assembly Generation with Auto-Regressive Transformer}

\author{Jingwen Ye}
\authornotemark[1]
\affiliation{%
 \institution{Tencent AIPD}
 \city{Shenzhen}
 \country{China}}
\email{jingwenye@tencent.com}
\author{Yuze He}
\authornote{Equal contributions.}
\affiliation{%
 \institution{Tsinghua University and Tencent AIPD}
 \city{Beijing}
 \country{China}
}
\email{hyz22@mails.tsinghua.edu.cn}
\author{Yanning Zhou}
\email{amandayzhou@tencent.com}
\authornotemark[2]
\author{Yiqin Zhu \thanks{}}
\author{Kaiwen Xiao}
\affiliation{%
 \institution{Tencent AIPD}
 \city{Shenzhen}
 \country{China}}
\author{Yong-Jin Liu}
\email{liuyongjin@tsinghua.edu.cn}
\authornotemark[2]
\affiliation{%
 \institution{Tsinghua University}
 \city{Beijing}
 \country{China}
}
\author{Wei Yang}
\author{Xiao Han}
\email{elehanx@gmail.com}
\authornote{Corresponding authors.}
\affiliation{%
 \institution{Tencent AIPD}
 \city{Shenzhen}
 \country{China}}

\begin{abstract}
Shape primitive abstraction, which decomposes complex 3D shapes into simple geometric elements, plays a crucial role in human visual cognition and has broad applications in computer vision and graphics. While recent advances in 3D content generation have shown remarkable progress, existing primitive abstraction methods either rely on geometric optimization with limited semantic understanding or learn from small-scale, category-specific datasets, struggling to generalize across diverse shape categories.
We present PrimitiveAnything, a novel framework that reformulates shape primitive abstraction as a primitive assembly generation task.
PrimitiveAnything includes a shape-conditioned primitive transformer for auto-regressive generation and an ambiguity-free parameterization scheme to represent multiple types of primitives in a unified manner. 
The proposed framework directly learns the process of primitive assembly from large-scale human-crafted abstractions, enabling it to capture how humans decompose complex shapes into primitive elements. 
Through extensive experiments, we demonstrate that PrimitiveAnything can generate high-quality primitive assemblies that better align with human perception while maintaining geometric fidelity across diverse shape categories.
It benefits various 3D applications and shows potential for enabling primitive-based user-generated content (UGC) in games.
{\it Project page: \url{https://primitiveanything.github.io}}

\end{abstract}




\begin{teaserfigure}
    \centering
    \vspace{-2pt}
    \includegraphics[width=1\textwidth]{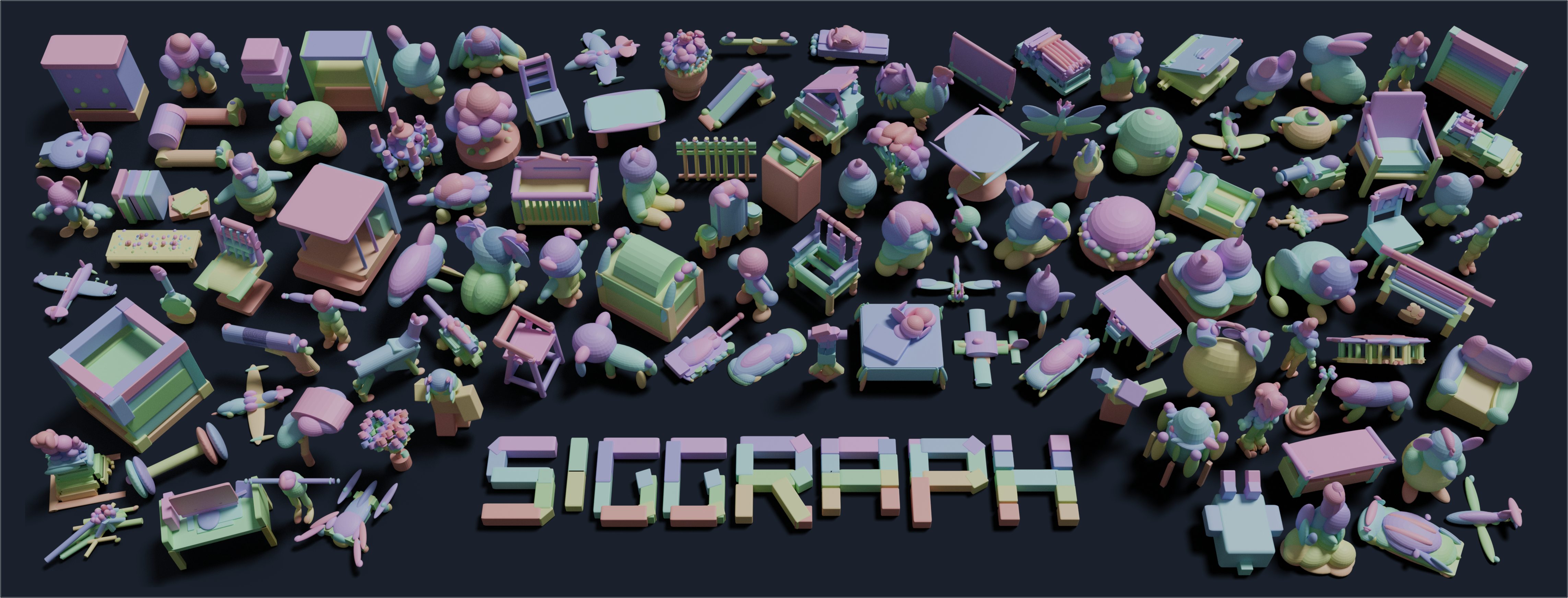}
    \caption{3D primitive assemblies created by PrimitiveAnything span diverse shape categories, enabling versatile primitive-based 3D content creation.}
    \label{fig:teaser}
    \vspace{7pt}
\end{teaserfigure}

\maketitle

\section{Introduction}

Understanding and representing 3D environments and objects has been a fundamental task in computer vision and graphics. 
Recent years have witnessed significant breakthroughs in 3D understanding and generation, with various representations including meshes~\cite{siddiqui2024meshgpt,chen2024meshanything,chen2024meshanythingv2}, point clouds~\cite{nichol2022point,vahdat2022lion}, and neural fields~\cite{poole2023dreamfusion,jun2023shap, hong2024lrm} enabling rapid generation of high-quality 3D contents. 
However, these representations, while effective for visualization and rendering, often lack the semantic structure and interpretability that align with human cognitive processes. 
Cognitive science research has long established that the human visual system possesses a remarkable ability to decompose complex visual scenes into simple geometric primitives - a process known as perceptual organization or shape abstraction~\cite{Biederman_1985, Biederman_2005}. 
This cognitive mechanism not only enables efficient visual processing and understanding but also facilitates our ability to reason about object structure, function, and physical interactions. 

Inspired by this human cognitive capability, the task of shape primitive abstraction and generation seeks to develop computational methods that can similarly decompose complex 3D shapes into interpretable primitive elements~\cite{roberts1963machine,binford1975visual}. 
This ability is not just theoretically interesting - it enables crucial applications in robotic manipulation, scene understanding, computer-aided design, and interactive modeling systems, where high-level structural understanding is essential for downstream tasks.

The development of primitive-based shape abstraction involves two fundamental design choices: the definition of primitive types to be used, and the computational approach to extract these primitives from raw 3D data. 
The selection of primitive types has evolved significantly over the past years. 
Early approaches primarily focused on simple geometric primitives such as cuboids~\cite{gupta2010blocks,tulsiani2017learning}, geons~\cite{Biederman_1985} and cylinders~\cite{binford1975visual}, which offer computational simplicity and intuitive interpretation but limited expressiveness.
Later work~\cite{pentland1986parts,Paschalidou2019CVPR} introduced superquadrics, which can represent a broader range of smooth shapes through parametric equations that generalize quadric surfaces, providing a better balance between representational power and computational tractability. 

For the extraction of primitives, technical approaches can be broadly categorized into two main streams: optimization-based methods and learning-based methods.
The former one formulates primitive detection as a geometric fitting problem, attempting to minimize various distance metrics between the primitive representations and input geometry~\cite{leonardis1997superquadrics,chevalier2003segmentation,liu2022robust}. 
These methods, while mathematically principled and interpretable, primarily focus on minimizing geometric surface distance between the original shape and primitive assemblies, with limited consideration of human abstraction logic. 
This often results in over-segmentation of semantic parts and fails to capture meaningful structural decomposition of shapes.
The latter approaches aim to learn primitive decomposition directly from data~\cite{zou20173d,tulsiani2017learning,Paschalidou2019CVPR,huang2023learning}.
However, these learning-based methods are typically trained on small-scale, category-specific datasets, leading to limited generalization capabilities across object categories. 
How to effectively parameterize primitives and learn generalizable abstraction concepts across diverse categories remains an open challenge.

Recent advances in 3D content generation~\cite{chen2024meshanythingv2,nichol2022point,hong2024lrm,zhang2024clay} have demonstrated the remarkable potential of directly learning the 3D representation from the large-scale 3D datasets, e.g. Objaverse~\cite{deitke2023objaverse}.
MeshAnything~\cite{chen2024meshanything,chen2024meshanythingv2}'s success in using an autoregressive transformer to generate human-crafted meshes that capture both geometric details and artistic intent.
Drawing on this insight, we reformulate primitive abstraction as a generation task, moving away from traditional geometric fitting or direct regression approaches.
Unlike previous methods that rely on hand-crafted optimization objectives or direct regression of primitive parameters, our generation-based framework learns to sequentially construct primitive abstractions in a manner similar to how humans might build up complex shapes from simple components.
This fundamentally different approach allows our method to better capture the hierarchical and semantic nature of shape decomposition while maintaining geometric accuracy.

The overall design follows two core concepts:
First, the \textit{primitive representation} must achieve \textbf{high geometric fidelity while compact enough} for efficient learning.
To this end, we utilize multiple types of primitives to jointly represent 3D shapes under a unified parameterization scheme. 
To address the inherent ambiguity in such parameterization and ensure stable training, we develop a comprehensive set of rules that uniquely define the parameter ordering and relationships between atomic elements, resulting in well-structured sequences suitable for learning.
Second, the \textit{learning framework} must possess \textbf{strong capacity} to handle complex shapes with varying numbers of primitives while remaining \textbf{primitive-agnostic} for extensibility. 
We address this through a shape-conditioned decoder-only transformer architecture that can generate variable-length primitive sequences. 
The framework's modular design treats primitive types as learnable tokens, enabling seamless integration of new primitive types without architectural changes, making it adaptable to different primitive representations.

Our main contribution can be summarized as follows:
1) We propose \textbf{PrimitiveAnything}, a novel primitive generation framework that reformulates shape abstraction as a sequence generation task, enabling the model to learn from and reproduce human-crafted shape decompositions.
2) We extend the single primitive representation to multiple primitives and design an ambiguity-free parameterization scheme, achieving high geometric fidelity while maintaining computational efficiency for learning.
3) PrimitiveAnything contains a shape-conditioned decoder-only transformer architecture that can handle variable-length primitive sequences and is easily extensible to new primitive types.
4) We demonstrate through extensive experiments that our method can generate high-quality primitive abstractions that better align with human perception compared to existing approaches, while maintaining geometric fidelity to the original shapes.

\section{Related Works}

\subsection{3D Content Generation}

Recent years have witnessed remarkable progress in 3D content generation, spanning diverse tasks from object generation~\cite{zhang2024clay,petrov2024gem3d,zhao2023michelangelo,dong2024coin3d} to texture synthesis~\cite{yu2024texgen,bensadoun2024meta,zhang2024mapa,hu2024diffusion,guerrero2024texsliders}. 
DreamFusion~\cite{poole2023dreamfusion} and SJC~\cite{wang2023score} pioneered the lifting of 2D diffusion models to 3D generation by optimizing neural radiance fields through score distillation sampling, with subsequent works like Magic3D~\cite{lin2023magic3d} and VSD~\cite{wang2023prolificdreamer} further refining this approach. 
The field has seen a shift towards data-driven large reconstruction models, starting with LRM~\cite{hong2024lrm} which leveraged transformers to generate triplane features from single images. 
This approach has been extended to handle multi-view inputs~\cite{li2024instantd,xu2024instantmesh,wang2024crm} and more efficient 3D representations~\cite{gslrm2024,TripoSR2024,yang2024tencent}, significantly improving generation fidelity.
Concurrent development of native 3D shape generation models has also shown promising results~\cite{zhao2023michelangelo, hui2024make, li2024craftsman,zhang20243d}. 
Notably, CLAY~\cite{zhang2024clay} introduced a two-stage approach combining a multi-resolution 3D shape VAE (extended from 3DShape2VecSet~\cite{zhang20233dshape2vecset}) with a DiT-based diffusion model~\cite{Peebles2022DiT} for high-quality shape generation.

These methods have explored various 3D representations, including point clouds~\cite{nichol2022point,vahdat2022lion}, meshes~\cite{siddiqui2024meshgpt,chen2024meshanything,chen2024meshanythingv2}, and neural fields~\cite{poole2023dreamfusion,jun2023shap, hong2024lrm}. 
However, while these representations excel at visualization and rendering, they typically lack the semantic abstraction and interpretability that align with human cognition.
Moreover, these generated meshes pose challenges for real-time multiplayer game environments, requiring both significant bandwidth for multiple users to load new content and additional optimization steps to meet game engine performance requirements.

\subsection{Shape Primitive Abstraction}
Shape primitive abstraction aims to represent 3D contents by "simple geometry shape", named as \textit{primitives}.
Prior approaches have used simple geometric primitives such as cuboids~\cite{gupta2010blocks,tulsiani2017learning,li2017grass,mo2019structurenet} and cylinders~\cite{binford1975visual}, which offer computational simplicity and intuitive interpretation but limited expressiveness. 
Super-quadrics~\cite{pentland1986parts,Paschalidou2019CVPR} provide a better balance between representational power and computational tractability by generalizing quadric surfaces.
Some methods~\cite{chen2020bsp,deng2020cvxnet}~proposed convex polytopes as primitive representations, offering different trade-offs between expressiveness and optimization complexity. 
Implicit primitives representing shapes through learned fields have also been explored~\cite{gadelha2020learning,genova2019learning,chen2019bae}.
Another line of works focus on Computer-aided design (CAD) modeling and define special primitives of Constructive Solid Geometry (CSG) trees~\cite{li2019supervised,le2021cpfn,li2023surface} via iterative boolean operators, which is beyond the scope of the paper.

To conduct shape primitive abstraction, optimization-based methods directly minimize reconstruction objectives, either through 3D supervision~\cite{liu2022robust,liu2023marching} to ensure geometric accuracy, or 2D supervision~\cite{monnier2023differentiable,gao2024learning} from multi-view images.
To overcome the local optima issues, EMS~\cite{liu2022robust} models the superquadric primitive abstraction probabilistically, enhancing its robustness to outliers.
However, these methods often fragment semantic parts into multiple pieces, as they primarily optimize for geometric reconstruction rather than human-like abstraction—a limitation stemming from the lack of large-scale datasets that capture human cognitive principles in shape decomposition.

Some works attempt to learn the shape distribution from data.
Pioneer work~\cite{tulsiani2017learning} presents a learning framework to assemble objects by predicting cuboid parameters, which was later extended to superquadrics by~\cite{Paschalidou2019CVPR}. To model the step-by-step construction process, 3D-PRNN~\cite{zou20173d} leveraged recursive neural networks (RNN) for sequential cuboid prediction. 
Recent work PASTA~\cite{li2024pasta} employs a sequence-to-sequence model for part-aware 3D shape generation. 
However, it utilizes only cuboids as primitives and trains exclusively on category-specific data, limiting its geometric expressiveness and generalization capabilities across different shape categories. 
Similarly, other learning methods also rely on small, category-specific datasets for training, constraining their applicability to diverse shape domains.

\subsection{Auto-Regressive Model for 3D Generation}

Auto-Regressive (AR) transformers have demonstrated impressive results on various tasks including language-modeling~\cite{radford2019language,brown2020language,achiam2023gpt,touvron2023llama} and vision generation~\cite{esser2021taming,ramesh2021zero,tian2024visual}. 
The core of AR models lies in their self-supervised learning strategy of predicting the next token in a sequence—a simple yet remarkably scalable and generalizable approach.

Due to their natural ability to handle variable-length outputs, AR models have been successfully applied to layout generation tasks. Sceneformer~\cite{wang2021sceneformer} pioneered this direction by introducing a transformer-based architecture to predict both categorical and geometric attributes of 3D objects for scene synthesis. 
This was followed by~\cite{paschalidou2021atiss, zhao2024roomdesigner} that further improved scene generation through better object sequence modeling and shape prior integration.

Recent works have demonstrated the potential of AR models in 3D artist-created mesh generation. 
MeshGPT~\cite{siddiqui2024meshgpt} first introduced the paradigm of treating meshes as sequences of vertices and faces. 
Building on this foundation, subsequent works achieved significant improvements through various innovations: introducing shape conditioning~\cite{chen2024meshanything}, developing more compact tokenization schemes~\cite{chen2024meshanythingv2,tang2024edgerunner}, and incorporating language capabilities~\cite{wang2024llamameshunifying3dmesh}. 
Particularly inspiring for our work is MeshAnything's~\cite{chen2024meshanything,chen2024meshanythingv2} approach to conditional mesh generation from point clouds, which motivates us to reformulate shape abstraction as a shape-conditioned generation task. 
We parameterize primitives as tokens and employ an auto-regressive model to predict the primitive sequence, effectively learning the implicit rules of shape decomposition.

\section{Method}

\begin{figure*}[t]
\centering
\includegraphics[width=1\linewidth]{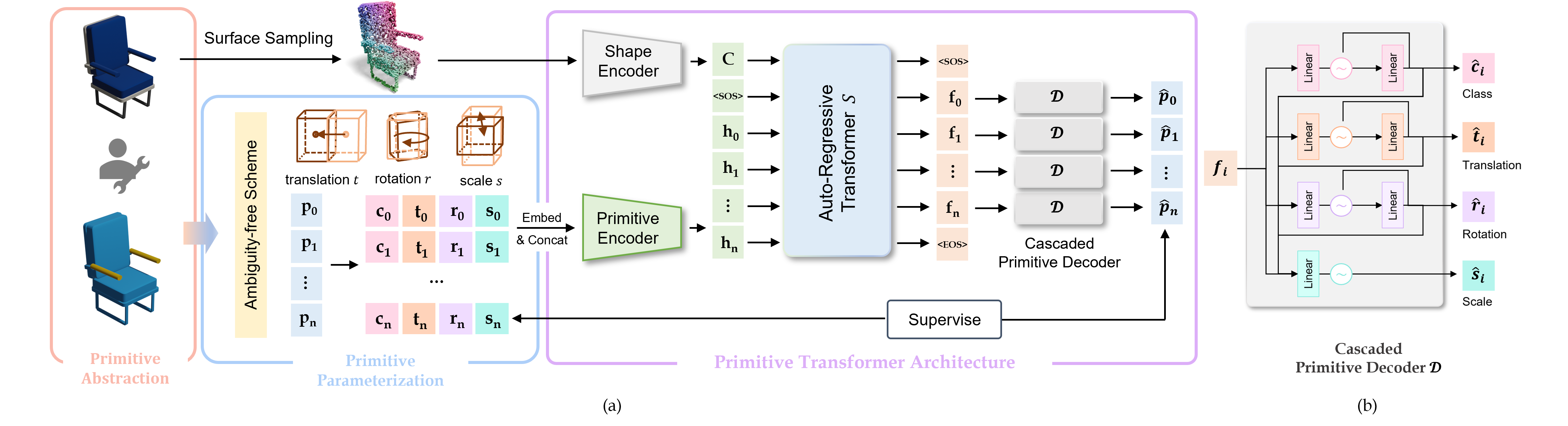}
\caption{Overview. We propose PrimitiveAnything to decompose complex shapes into 3D primitive assembly via the auto-regressive transformer. 
Given human-crafted 3D primitive abstraction contents, we first design an ambiguity-free scheme to parameterize each primitive $p$ into class label $c$, translation $t$, rotation $r$ and scale $s$, and then employ a primitive encoder to form primitive token $h$. 
Meanwhile, a shape encoder encodes 3D shape features $\mathcal{C}$ from sampled point clouds.
Our primitive transformer $\mathcal{S}$ predicts the next primitive based on
the input condition $\mathcal{C}$ and previously generated primitives. To model the dependencies among primitive attributes, we proposed a
cascaded primitive decoder $\mathcal{D}$ that sequentially predicts
primitive attributes.
}
\label{fig:archi}
\end{figure*}
Our proposed \textbf{PrimitiveAnything} is a novel primitive generation framework that reformulates shape abstraction as a sequence generation task, enabling human-like shape decomposition. 
Our method comprises three key components: an ambiguity-free primitive parameterization scheme (Sec.~\ref{subsec:param}), a primitive transformer architecture (Sec.~\ref{subsec:network}), and an auto-regressive generation pipeline (Sec.~\ref{subsec:gen}). Fig.~\ref{fig:archi} illustrates the overall framework.

\subsection{Primitive Parameterization}
\label{subsec:param}

Our goal is to establish a parameterization scheme that represents 3D objects using an arbitrary number and variety of predefined primitives. 
Given a 3D object and a predefined 3D primitive set $\mathcal{P} = \{ \mathcal{P}_1, \ldots, \mathcal{P}_N \}$ with $N$ standard primitive shapes, we aim to find its approximate representation $\hat{\mathcal{M}}$ composed of $n$ transformed primitives:
\begin{align}
    \hat{\mathcal{M}} & = \{ p_1,p_2,\dots,p_n \}
\end{align}
Each primitive $p_i$ is defined by combining a standard primitive type $\mathcal{P}_{c_i} \in \mathcal{P}$ with its rigid transformation $\mathcal{T}(\cdot)$ in 3D space:
\begin{align}
    p_i & = \mathcal{T}(\mathcal{P}_{c_i}; \mathbf{s}_i, \mathbf{r}_i, \mathbf{t}_i)
\label{eq:prim}
\end{align}
where:
\begin{itemize}
    \item $c_i$ denotes the primitive class label
    \item $\mathbf{s}_i \in \mathbb{R}^3$ denotes the scale factors along three principal axes
    \item $\mathbf{r}_i \in \mathbb{R}^3$ specifies the rotation angles in x-y-z Euler order
    \item $\mathbf{t}_i \in \mathbb{R}^3$ defines the translation of the primitive center
\end{itemize}
These transformation components are sequentially applied to the standard predefined primitive to achieve the final configuration $p_i$.

\begin{figure}[!t]
\centering
\includegraphics[width=0.9\linewidth]{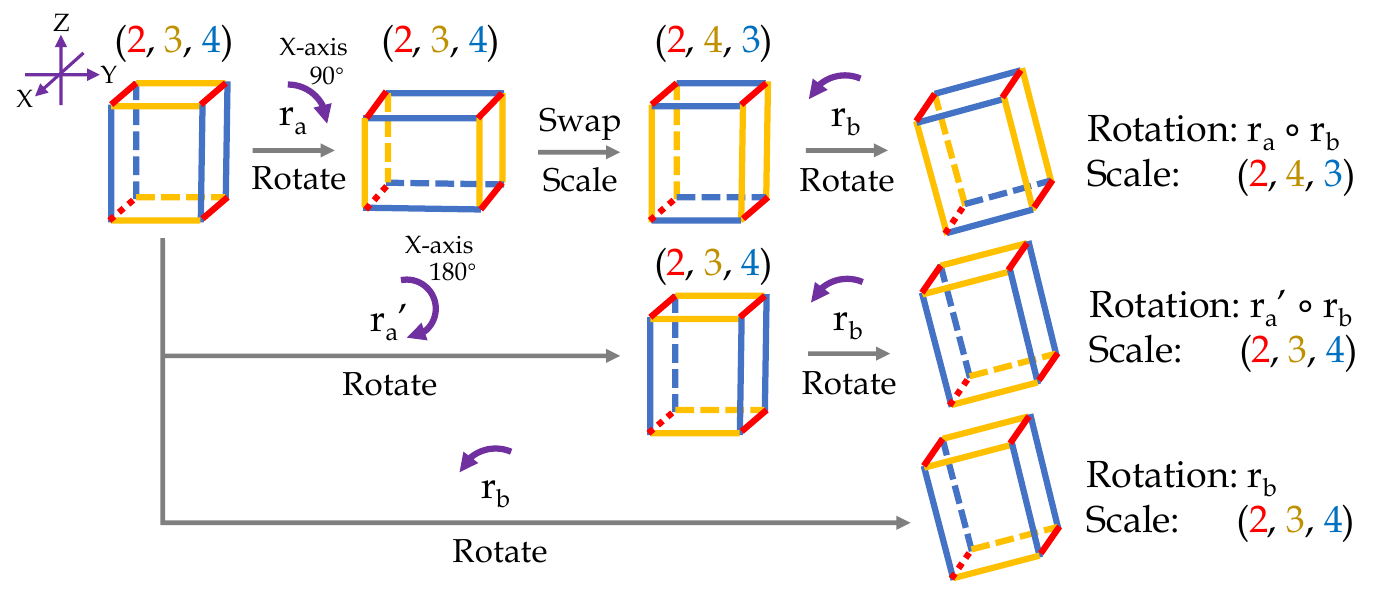}
\caption{Demonstration of primitive attribute ambiguity. A primitive with inherent symmetry can correspond to multiple scales and rotations through self-rotation and possible axis swapping.}
\vspace{-10pt}
\label{fig:symmetry}
\end{figure}

However, directly using the above parameterization is not sufficient.
Upon deeper analysis, we observe that many common primitives (such as cuboids and cylinders) possess inherent symmetries. 
Due to these symmetries, different combinations of scale and rotation parameters can produce identical transformed primitives in 3D space, creating ambiguity in the parameter representation.
Consider the cuboid example in Fig.~\ref{fig:symmetry}: applying a 90-degree rotation around the z-axis before the original rotation, while simultaneously swapping the x and y scale factors, results in identical shapes. 
Such parameter ambiguity complicates the learning process, as the model encounters multiple valid parameter combinations for the same shape, which cannot be resolved through mathematical reformulation alone.

To address this issue, we propose an ambiguity-free parameterization approach. 
Let $V$ denote the set of symmetry axes for the predefined primitive $\mathcal{P}_{c_i}$ corresponding to the transformed primitive $p_i$, $m_j$ represents the order of symmetry for the $j$-th symmetry axis $\mathbf{v}_j \in V$.
Note that axis permutations that result in equivalent configurations are also counted when determining the total symmetry order.
We can then define rotational symmetry set $\mathcal{R}$ as:
\begin{align}
    \mathcal{R} = \bigcup_{j=1}^n \left\{ \text{Rot}(\mathbf{v}_j, \frac{2\pi k}{m_j}) \,\middle|\, k = 0, 1, \dots, m_j - 1 \right\}
\end{align}
where $\text{Rot}(\mathbf{v},\theta)$ represents a rotation transformation of angle $\theta$ around axis $\mathbf{v}$. 
We further compose each equivalent rotation transformation $\mathbf{r}_{k}\in \mathcal{R}$ with the original transformation parameters $(\mathbf{s}_i, \mathbf{r}_i, \mathbf{t}_i)$, and select the combination that yields the minimal L1 norm of rotation as our new transformation $(\mathbf{s}_i', \mathbf{r}_i', \mathbf{t}_i')$:
\begin{align}
    & \mathbf{r}_i' = \arg\min_{\mathbf{r}_k \in \mathcal{R}} \|\hat{\mathbf{r}}_k\|_1,\quad \text{where} \\
& \mathcal{T}(\cdot; \hat{\mathbf{s}}_k, \hat{\mathbf{r}}_k, \hat{\mathbf{t}}_k) = \mathcal{T}(\mathcal{T}(\cdot;\mathbf{s}_k,\mathbf{r}_k); \mathbf{s}_i, \mathbf{r}_i, \mathbf{t}_i)
\end{align}
Consequently, we reformulate the transformed primitive as:
\begin{align}
    p_i & = \mathcal{T}(\mathcal{P}_{c_i}; \mathbf{s}_i', \mathbf{r}_i', \mathbf{t}_i')
\end{align}
This formulation eliminates symmetry-induced ambiguity while reducing the parameter space, facilitating more effective learning and preventing mode confusion.

\subsection{Primitive Transformer}
\label{subsec:network}
Drawing inspiration from how humans sequentially compose shapes by assembling basic geometric elements, we formulate primitive abstraction as a sequential generation process. 
Our primitive transformer $F$ predicts the next primitive based on the input condition $\mathcal{C}$ and previously generated primitives:
\begin{align}
    p_i=F(\mathcal{C};p_1,\dots,p_{i-1})
\end{align}
The architecture consists of three learnable modules: a primitive encoder $\mathcal{E}$, a decoder-only transformer model $\mathcal{S}$, and a cascaded primitive decoder $\mathcal{D}$. 
We discretize the scale, rotation, and translation parameters, and treat them along with the class label as discrete input tokens.
For the $i$-th primitive $p_i$, its attributes $c_i,\mathbf{s}_i,\mathbf{r}_i,\mathbf{t}_i$ are embedded using the learnable embeddings $\mathbf{e}_c$, $\mathbf{e}_s$, $\mathbf{e}_r$, $\mathbf{e}_t$, followed by a linear primitive encoder $\mathcal{E}$ to generate the primitive token $\mathbf{h}_i$:  
\begin{align}
    \mathbf{h}_i = \mathcal{E}(\mathbf{e}_c(c_i), \mathbf{e}_s(\mathbf{s}_i), \mathbf{e}_r(\mathbf{r}_i), \mathbf{e}_t(\mathbf{t}_i))
\end{align}
The decoder-only transformer $\mathcal{S}$ receives the input condition $\mathcal{C}$ and tokens of all previously generated primitives as input, producing the feature representation $\mathbf{f}_i$ of the new primitive:
\begin{align}
    \mathbf{f}_i = \mathcal{S}(\mathcal{C}; \mathbf{h}_1,\dots,\mathbf{h}_{i-1})
\end{align}
To generate the next primitive's attributes from $\mathbf{f}_i$, similar to previous scene generation works~\cite{ritchie2019fast,wang2021sceneformer,paschalidou2021atiss}, we utilize a cascaded primitive decoder that explicitly models the dependencies among primitive attributes:
\begin{align}
\hat{c}_i &= \mathcal{D}_c(\mathbf{f}_i) \\
\hat{\mathbf{t}}_i &= \mathcal{D}_t(\mathbf{f}_i, \mathbf{e}_c(c_i)) \\
\hat{\mathbf{r}}_i &= \mathcal{D}_r(\mathbf{f}_i, \mathbf{e}_c(c_i), \mathbf{e}_t(t_i)) \\
\hat{\mathbf{s}}_i &= \mathcal{D}_s(\mathbf{f}_i, \mathbf{e}_c(c_i), \mathbf{e}_t(t_i), \mathbf{e}_r(r_i))
\end{align}
where $\mathcal{D}_c$, $\mathcal{D}_t$, $\mathcal{D}_r$, and $\mathcal{D}_s$ represent the class, translation, rotation, and scale decoders respectively. 
Each decoder takes the concatenation of the initial feature $\mathbf{f}_i$ and the embedded representations of previously decoded attributes, and then outputs logits of the probability.
This design captures the natural correlations between primitive attributes: the choice of primitive type influences its likely position, rotation, and scale parameters, and also aligns with human assembling logic: selecting type, determining position, and then adjusting rotation and scale.

\subsection{Auto-Regressive Primitive Generation}
\label{subsec:gen}
\xhdr{Sequence Formulation}
Our primitive transformer can be trained for shape-conditioned generation by taking condition features before primitive features through the carefully designed framework.
We select point clouds as input conditions, leveraging their ease of extraction from various 3D representations, and utilize the Michelangelo~\cite{zhao2023michelangelo} encoder to convert the point cloud into a fixed-length token sequence. 
This encoded sequence is concatenated with a start token \text{<SOS>}, followed by the primitive tokens $\{h_1,...,h_{i-1}\}$, forming the complete input sequence for the transformer. 
To determine when generation should terminate, we introduce an \text{<EOS>} decoder $\mathcal{D}_{eos}$ operating on the primitive feature $\mathbf{f}_i$ output by the transformer.
Primitives are sorted by centroids in z-y-x order (z-axis as top), progressing from lowest to highest.

\xhdr{Training objective}
We train the primitive transformer using next-step prediction as the primary objective, while incorporating an auxiliary 3D shape guidance term:
\begin{align}
    \mathcal{L}=\mathcal{L}_{eos} + \mathcal{L}_{ce}+\mathcal{L}_{cd}
\end{align}
Here, $\mathcal{L}_{ce}$ denotes the cross-entropy loss used to supervise the discrete primitive attributes $c_i, \mathbf{s}_i, \mathbf{r}_i, \mathbf{t}_i$, while $\mathcal{L}_{eos}$ represents the binary cross-entropy loss applied to $\mathcal{D}_{eos}(\mathbf{f}_i)$ to guide the termination prediction. 
To ensure precise alignment and robust control over reconstruction quality, the Chamfer Distance loss~\cite{fan2017point} $\mathcal{L}_{cd}$ is employed for each generated primitive. 
As the predicted primitive attributes are discrete, the Gumbel-Softmax technique~\cite{jang2017categorical} is applied to enable differentiable sampling for each next-token prediction, generates the predicted attributes $\{\mathbf{s}_{i,\text{pred}}\}_{i=1}^n$, $\{\mathbf{r}_{i,\text{pred}}\}_{i=1}^n$, and $\{\mathbf{t}_{i,\text{pred}}\}_{i=1}^n$, forming the predicted next-primitive set $\{p_{i,\text{pred}}\}_{i=1}^n$. 
Subsequently, both $\{p_{i,\text{pred}}\}_{i=1}^n$ and the ground-truth primitive set $\{p_{i,\text{gt}}\}_{i=1}^n$ are sampled to produce the point clouds $pc_{\text{pred}}$ and $pc_{\text{gt}}$, respectively. The Chamfer Distance loss is then calculated as: 
\begin{align}
    \mathcal{L}_{cd} = CD(pc_{\text{pred}}, pc_{\text{gt}})
\end{align}
where $CD(\cdot, \cdot)$ denotes the Chamfer distance \cite{fan2017point}.

\xhdr{Inference}
Starting from an input point cloud, our primitive transformer autoregressively generates primitive features $\{\mathbf{f}_i\}_{i=1}^n$, which are subsequently decoded and assembled into the final primitive representation $\mathcal{\hat{M}}$.
This process continues until the \text{EOS} judgment, signaling the completion of the primitive generation.  

\subsection{Implementation Details}
For model architecture, our auto-regressive transformer has 12 layers with a hidden size of 768. 
The cascaded decoders are implemented as 2-layer MLPs (hidden size 768) that process the concatenation of primitive features and previously decoded attribute embeddings.
All training data was normalized to lie within a unit cube.
For primitive attribute discretization, rotation, scale and translation are discretized into 180, 128, and 128 levels per dimension, respectively.
Attribute embeddings are 16-dimensional for rotation, scale, and translation parameters, with 48-dimensional embeddings for class labels.
The training was conducted using the Adam optimizer with a learning rate of $1 \times 10^{-3}$, a batch size of 128, and gradient accumulation over 4 steps. The model was trained on 8 NVIDIA V100 GPUs for 3 days.

\section{Experiments}

\subsection{Experimental Setup}
\label{subsec:setup}
\xhdr{Datasets}
We collect a large-scale 3D dataset with primitive abstraction annotations created by human annotators, named HumanPrim. 
HumanPrim contains 120K samples, each consisting of a 3D mesh, its surface point cloud, and manual primitive assembly.
Three primitive types are utilized in the assembly: cuboids, elliptical cylinders, and ellipsoids.
The primitive sequences have an average length of 30.9 primitives, with the longest sequence containing 144 primitives.
To ensure a thorough evaluation of our method, we randomly select 314 high quality samples with artist-created labels to form the test set.
To evaluate the proposed method's generalization capability, we additionally evaluate on data from ShapeNet~\cite{shapenet2015} and Objarverse ~\cite{deitke2023objaverse}.

\xhdr{Evaluation Metrics}
For geometric evaluation, we uniformly sample point clouds on the surfaces of predicted primitives and compare them against ground-truth point clouds sampled from the original meshes. We employ four metrics for evaluation: Chamfer Distance (CD), Earth Mover's Distance (EMD), Hausdorff Distance, and Voxel-IoU. For the Voxel-IoU metric, both predicted and ground-truth point clouds are voxelized at a resolution of $32^3$, after which their intersection over union is computed.
Moreover, to evaluate how well the generated primitive abstractions align with human decomposition patterns, we additionally employ instance segmentation metrics.
Following previous works~\cite{xie2022planarrecon,he2024alphatablets}, we address the geometric discrepancy between ground-truth and predicted primitives through a label transfer process: points are first sampled on the ground-truth mesh, then matched to their nearest neighbors in the predicted primitives to transfer prediction labels. Three segmentation metrics are used: rand index (RI), variation of information (VOI), and segmentation covering (SC).

\subsection{Comparisons}

\begin{figure*}[t]
\centering
\includegraphics[width=1.0\linewidth]{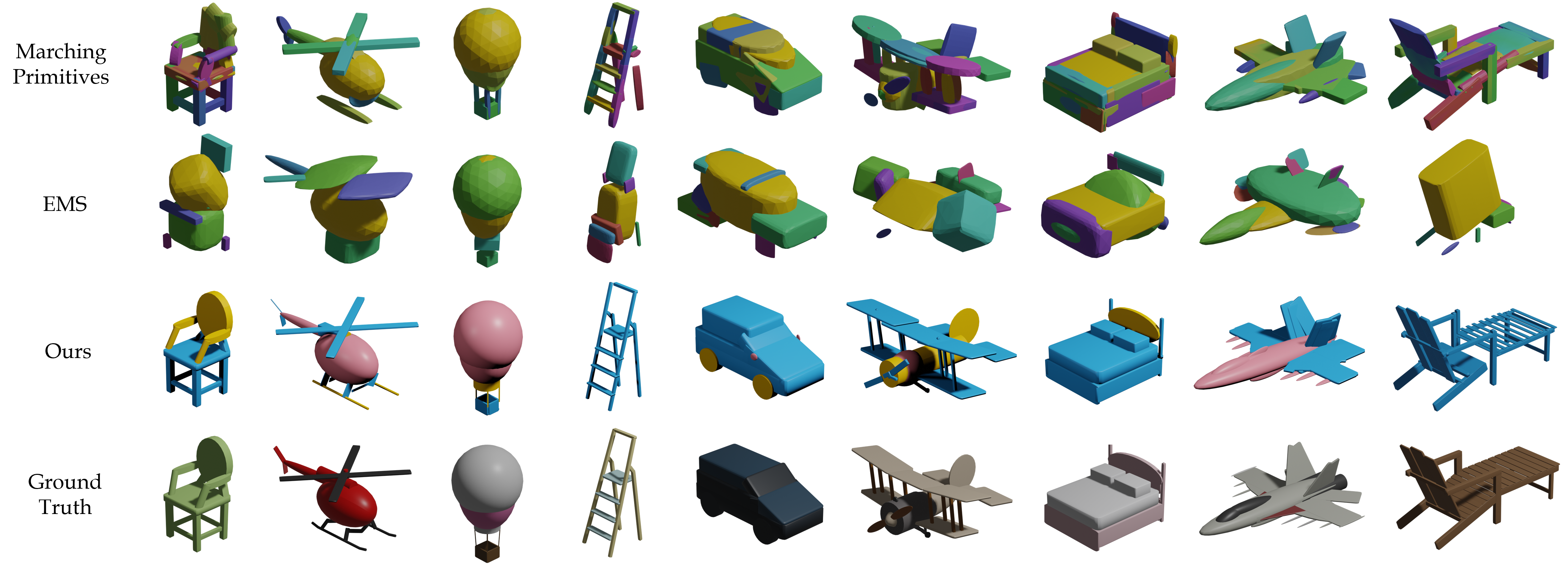}
\vspace{-18pt}
\caption{
Qualitative comparisons on the HumanPrim test set: In our method, colors indicate different primitive types, while in Marching Primitives and EMS, colors represent separate superquadrics. Our method achieves human-crafted primitive abstraction and faithfully reproduces the original 3D structure.}

\label{fig:main}
\end{figure*}

\xhdr{Comparison methods}
As no existing methods can generate variable-length sequences of diverse primitive types, we compare our approach with state-of-the-art optimization-based methods: Marching-Primitives (MP)~\cite{liu2023marching} and EMS~\cite{liu2022robust}, both using superquadric primitives.
We further compare against two learning-based methods: a cuboid-based approach~\cite{tulsiani2017learning} and a superquadric-based method~\cite{Paschalidou2019CVPR}, using their pre-trained models.
Note that these comparisons are limited to their specifically trained categories, as these methods do not generalize to other shape classes.

\begin{table}[htbp]
\centering
\caption{Geometric comparison with optimization-based methods on the HumanPrim test set.}
\vspace{-5pt}
\resizebox{1.0\linewidth}{!}{
\begin{tabular}{l|c|c|c|c}
\toprule
Method & CD~$\downarrow$ & EMD~$\downarrow$ & Hausdorff~$\downarrow$ & Voxel-IoU~$\uparrow$ \\
\midrule
EMS~\cite{liu2022robust} & 0.1062 &	0.0840 &	0.338 &	0.259 \\
MP~\cite{liu2023marching} & 0.0546 & 0.0515 & \textbf{0.120} & 0.201 \\
Ours & \textbf{0.0404} & \textbf{0.0475} & 0.158 & \textbf{0.484} \\
\bottomrule
\end{tabular}
}
\label{tab:geo}
\end{table}

\xhdr{Quantitative Comparisons}
We present quantitative comparisons between our method and optimization-based approaches in Tab.~\ref{tab:geo}. While EMS shows robustness in fitting single primitives, it faces challenges with multiple primitive predictions, particularly in identifying smaller primitive parts, and often fails in the absence of a main primitive. Marching-Primitives achieves progressively refined 3D contour matching through iterative optimization, as reflected in its Hausdorff distance performance (maximum distance among nearest point pairs between two point clouds). However, its results frequently deviate from human construction patterns, often decomposing regions that should be represented by a single primitive into multiple primitives. This generates erroneous occupancy within primitives, leading to notably lower Voxel-IoU scores, which measure surface coverage effectiveness, and reduced overall 3D accuracy as indicated by CD and EMD metrics. Our method demonstrates superior overall performance across metrics. The 3D instance segmentation metrics shown in Tab.~\ref{tab:seg} further validate our method's superior capability in generating human-like primitive abstractions.

\begin{table}[!t]
\centering
\caption{3D segmentation accuracy comparison with optimization-based methods on the HumanPrim test set.}
\vspace{-5pt}
\begin{tabular}{l|c|c|c}
\toprule
Method & RI~$\uparrow$ & VOI~$\downarrow$ & SC~$\uparrow$ \\
\midrule
EMS~\cite{liu2022robust} & 0.696 &	3.520 &	0.280 \\
MP~\cite{liu2023marching} & 0.821 & 3.793 & 0.254 \\
Ours & \textbf{0.892} & \textbf{2.296} & \textbf{0.409} \\
\bottomrule
\end{tabular}
\label{tab:seg}
\end{table}

We conduct additional comparisons with learning-based methods on the Chair subset of our HumanPrim test set (59 samples) and the {\it Chair} category from ShapeNet's test split (1,317 samples), as shown in Tab.~\ref{tab:geochair}, since previous learning-based approaches are limited to training on individual categories. ~\cite{tulsiani2017learning} predicts only cuboids and shows a limited ability to model objects effectively. ~\cite{Paschalidou2019CVPR} utilizes superquadrics and offers more flexible object modeling capabilities, but its overall accuracy remains insufficient.
Notably, our method demonstrates robust generalization, outperforming all benchmarked approaches across all metrics, even though it was not trained on the ShapeNet dataset—unlike the comparison methods, which were specifically designed for it.
This superiority is further corroborated by the segmentation metrics presented in Tab.~\ref{tab:segchair} (ShapeNet is not tested due to the absence of 
ground-truth primitive labels).

\begin{table}[!t]
\centering
\caption{Geometric comparison with learning-based methods on HumanPrim test set (Chair subset) and ShapeNet test set ({\it Chair} category).}
\vspace{-5pt}
\resizebox{1.0\linewidth}{!}{
\begin{tabular}{l|c|c|c|c}
\toprule
Method & CD~$\downarrow$ & EMD~$\downarrow$ & Hausdorff~$\downarrow$ & Voxel-IoU~$\uparrow$ \\
\midrule
\multicolumn{5}{c}{Chair subset of HumanPrim} \\
\midrule
\cite{tulsiani2017learning} & 0.2512 & 0.1835 & 0.420 & 0.041 \\
\cite{Paschalidou2019CVPR} & 0.1438 & 0.1088 & 0.332 & 0.095 \\
Ours & \textbf{0.0343} & \textbf{0.0458} & \textbf{0.136} & \textbf{0.550} 
\\
\midrule
\multicolumn{5}{c}{{\it Chair} category of ShapeNet} \\
\midrule
\cite{tulsiani2017learning} & 0.2282 &	0.1667 &	0.411 &	0.046 \\
\cite{Paschalidou2019CVPR} & 0.1343	& 0.1038 &	0.285 &	0.094 \\
Ours & \textbf{0.0553} &	\textbf{0.0588}	& \textbf{0.190} &	\textbf{0.322}
\\
\bottomrule
\end{tabular}
}
\label{tab:geochair}
\end{table}

\begin{table}[!t]
\centering
\caption{3D segmentation accuracy comparison with learning-based methods on the HumanPrim test set (Chair subset).}
\vspace{-5pt}
\begin{tabular}{l|c|c|c}
\toprule
Method & RI~$\uparrow$ & VOI~$\downarrow$ & SC~$\uparrow$ \\
\midrule
\cite{tulsiani2017learning} & 0.740 & 3.097 & 0.335 \\
\cite{Paschalidou2019CVPR} & 0.660 & 3.346 & 0.274 \\
Ours & \textbf{0.931} & \textbf{1.499} & \textbf{0.578} \\
\bottomrule
\end{tabular}
\vspace{-10pt}
\label{tab:segchair}
\end{table}

\xhdr{Qualitative Comparisons}
Fig.~\ref{fig:main} presents qualitative comparisons with optimization-based methods. EMS produces sparse and coarse superquadrics abstractions that lack detailed surface fidelity. Marching-Primitives achieves rough shape contours through highly overlapping primitives, its decompositions often deviate from human construction patterns. Specifically, it tends to over-segment large or elongated parts using numerous primitives and frequently overlooks fine structural details. In contrast, our method successfully identifies geometric features at various scales, achieving both human-crafted shape abstraction and faithful reproduction of the overall surfaces and global structure of the original 3D objects.

\begin{figure*}[htb]
\centering
\begin{minipage}[t]{0.48\textwidth}
    \centering
    \includegraphics[width=1\linewidth]{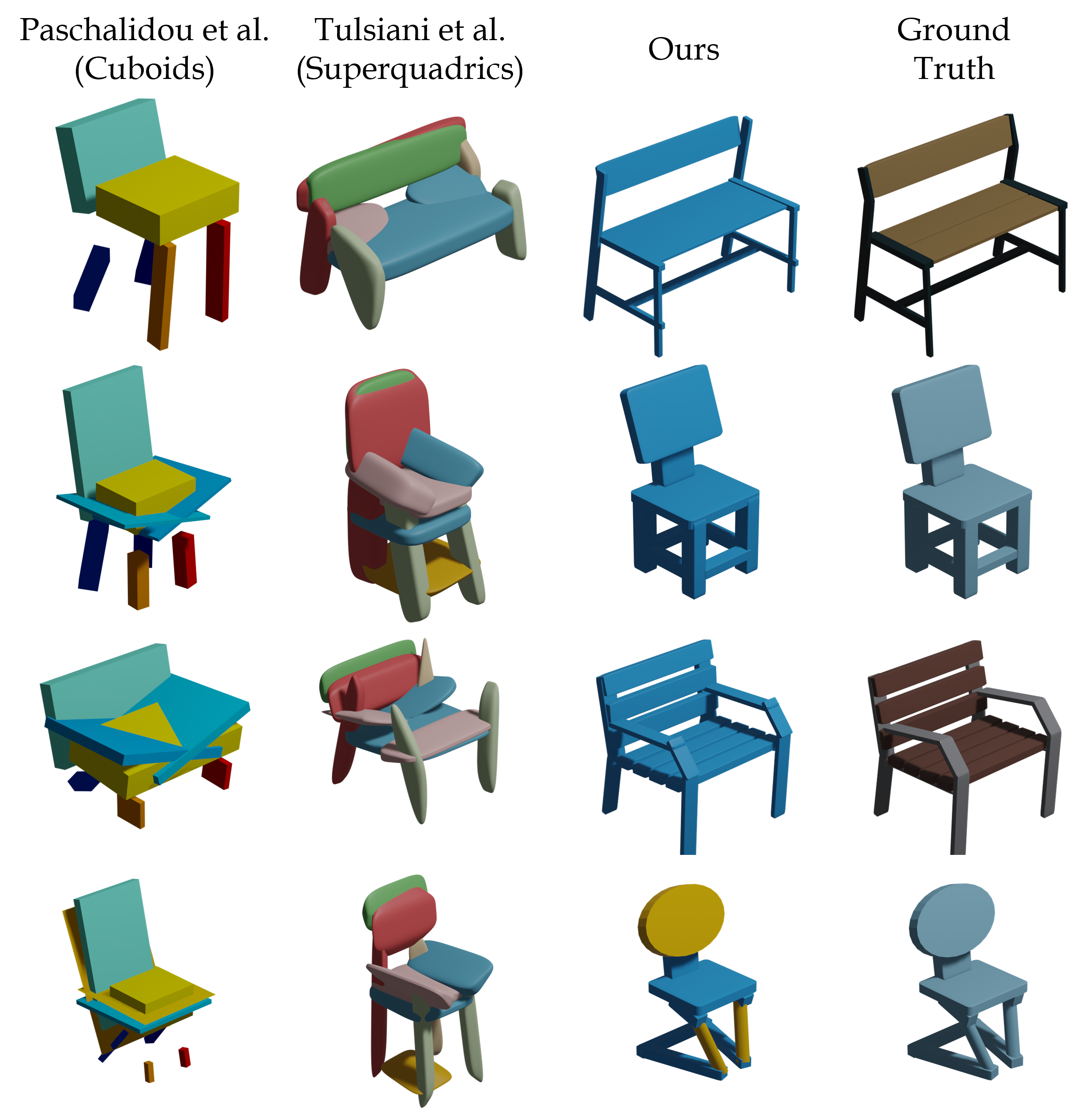}
    \caption{Comparisons on the HumanPrim test set (Chair subset).}
    \label{fig:chair}
\end{minipage}
\quad
\begin{minipage}[t]{0.49\textwidth}
    \centering
    \includegraphics[width=1\linewidth]{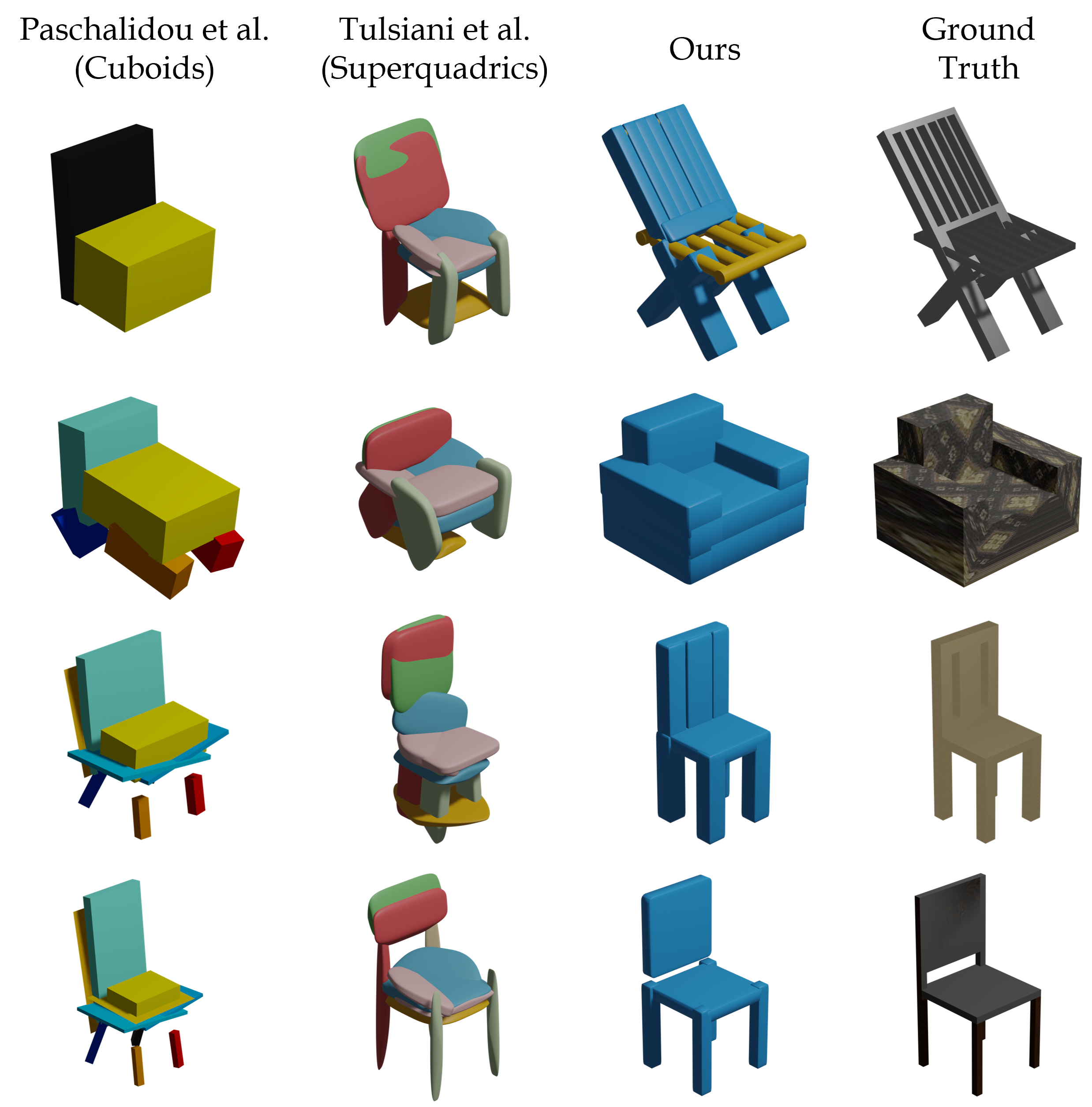}
    \caption{Comparisons on the ShapeNet test set (Chair category).}
    \label{fig:shapenet}

\end{minipage}
\end{figure*}

\begin{figure*}[t]
\centering
\includegraphics[width=0.93\linewidth]{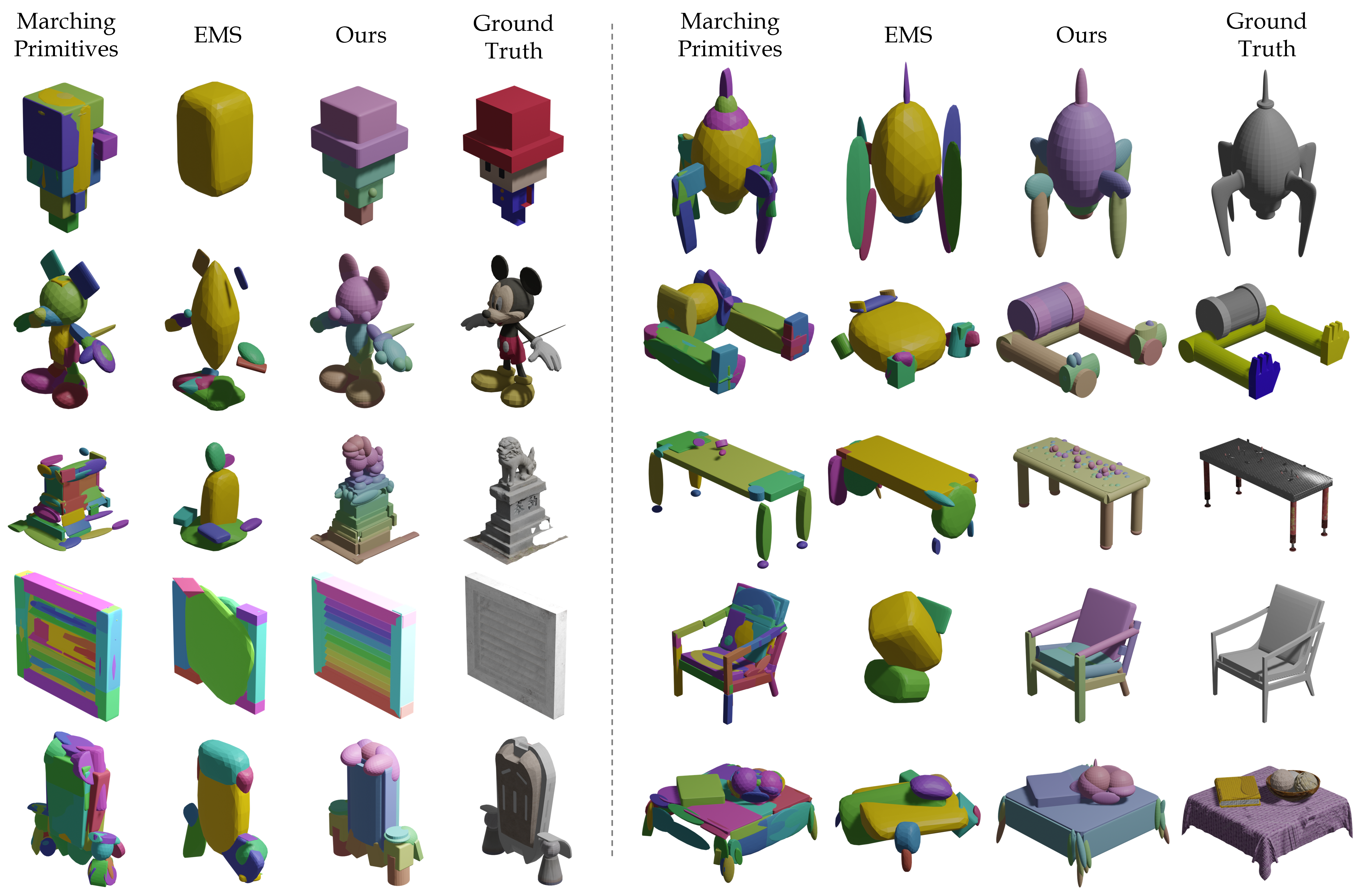}
\vspace{-2pt}
\caption{
Comparisons on the Objaverse dataset.
}
\label{fig:objaverse}
\end{figure*}

Figs.~\ref{fig:chair} and \ref{fig:shapenet} illustrate visual comparisons with other learning-based methods on the Chair subset. ~\cite{tulsiani2017learning} produces sparse cuboid abstractions with relatively coarse geometric structures. Although ~\cite{Paschalidou2019CVPR}'s multiple superquadric predictions better conform to 3D object surfaces, it still exhibits numerous imprecise and erroneous predictions. In contrast, our method demonstrates significant advantages in both accuracy and generalization capacity.
Fig.~\ref{fig:objaverse} provides qualitative comparisons on the Objaverse dataset, further demonstrating the generalizability of our method across diverse 3D objects.

\subsection{Ablation Study}

\begin{table}[!t]
\centering
\caption{Ablation studies on the HumanPrim test set.}
\vspace{-5pt}
\resizebox{1.0\linewidth}{!}{
\begin{tabular}{l|c|c|c|c}
\toprule
Method & CD~$\downarrow$ & EMD~$\downarrow$ & Hausdorff~$\downarrow$ & Voxel-IoU~$\uparrow$ \\
\midrule
w/o ambiguity-free param. & 0.0564 & 0.0584 & 0.204 & 0.414 \\
w/o cascaded decoding & 0.0558 & 0.0586 & 0.243 & 0.458 \\
w/o Chamfer Distance loss & 0.0440 & 0.0514 & 0.174 & 0.475 \\
Ours & \textbf{0.0404} & \textbf{0.0475} & \textbf{0.158} & \textbf{0.484} \\
\bottomrule
\end{tabular}
}
\vspace{-8pt}
\label{tab:abl}
\end{table}

To validate the effectiveness of each component in our framework, we conduct ablation studies using the HumanPrim dataset while keeping the experimental and training configurations consistent with those in Sec.~\ref{subsec:setup}. We sequentially disable specific modules while leaving others unchanged.

The results in Tab.~\ref{tab:abl} show that all proposed improvements contribute effectively to the overall performance. The ambiguity-free parameterization scheme helps reduce mode confusion, as evidenced by the Voxel-IoU metric. The cascaded decoding architecture improves generation stability and prevents outlier occurrences, as reflected in a decrease of the mean Hausdorff distance. The Chamfer Distance loss allows for finer-grained control over primitive generation, leading to improved accuracy and detail. These results show that each component of our method is essential for high-quality shape abstraction.

\begin{figure}[t]
\centering
\vspace{-5pt}
\includegraphics[width=1.0\linewidth]{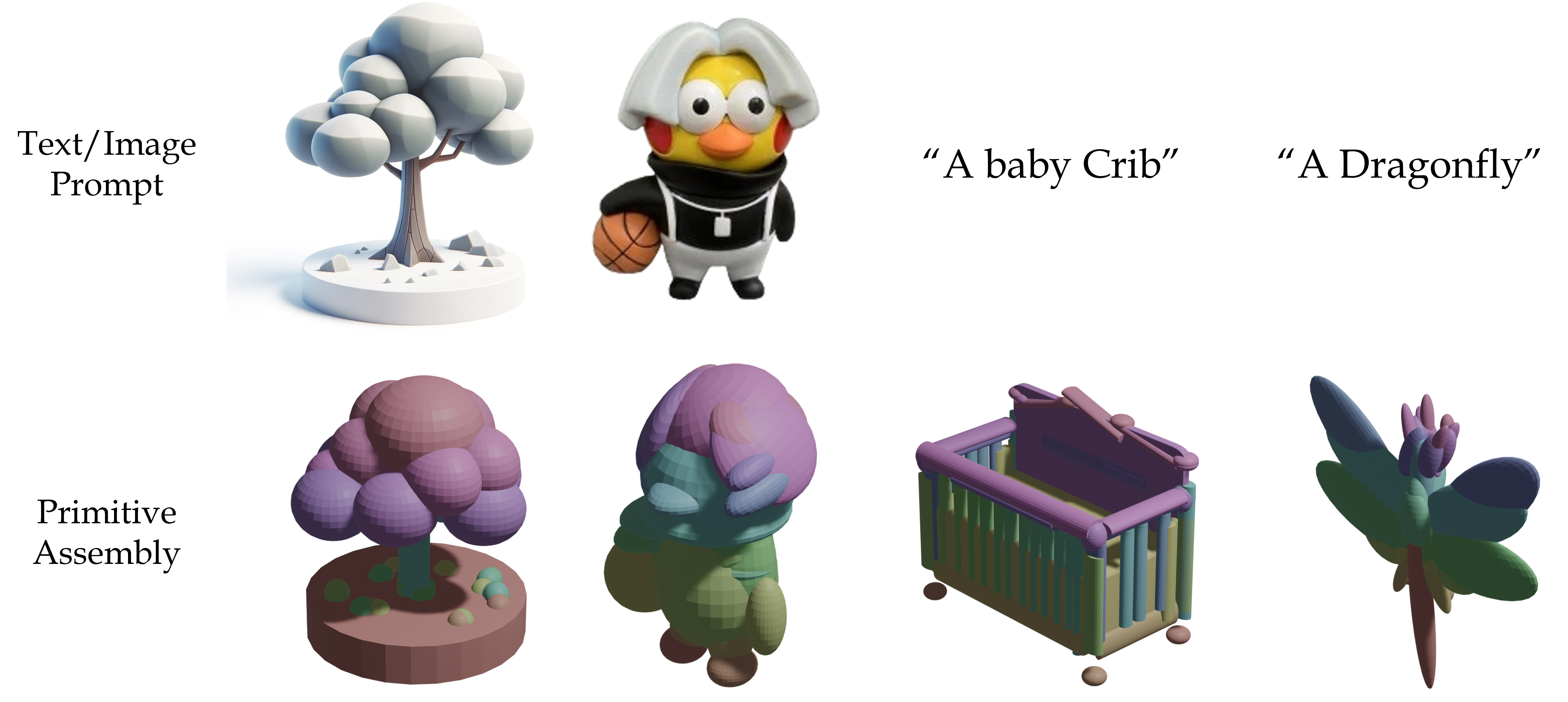}
\vspace{-15pt}
\caption{
PrimitiveAnything interfaces with state-of-the-art 3D shape generation models to enable text- and image-conditioned primitive-based 3D content generation.
}
\label{fig:app}
\vspace{-10pt}
\end{figure}


\begin{figure*}[t]
\centering
\includegraphics[width=0.98\linewidth]{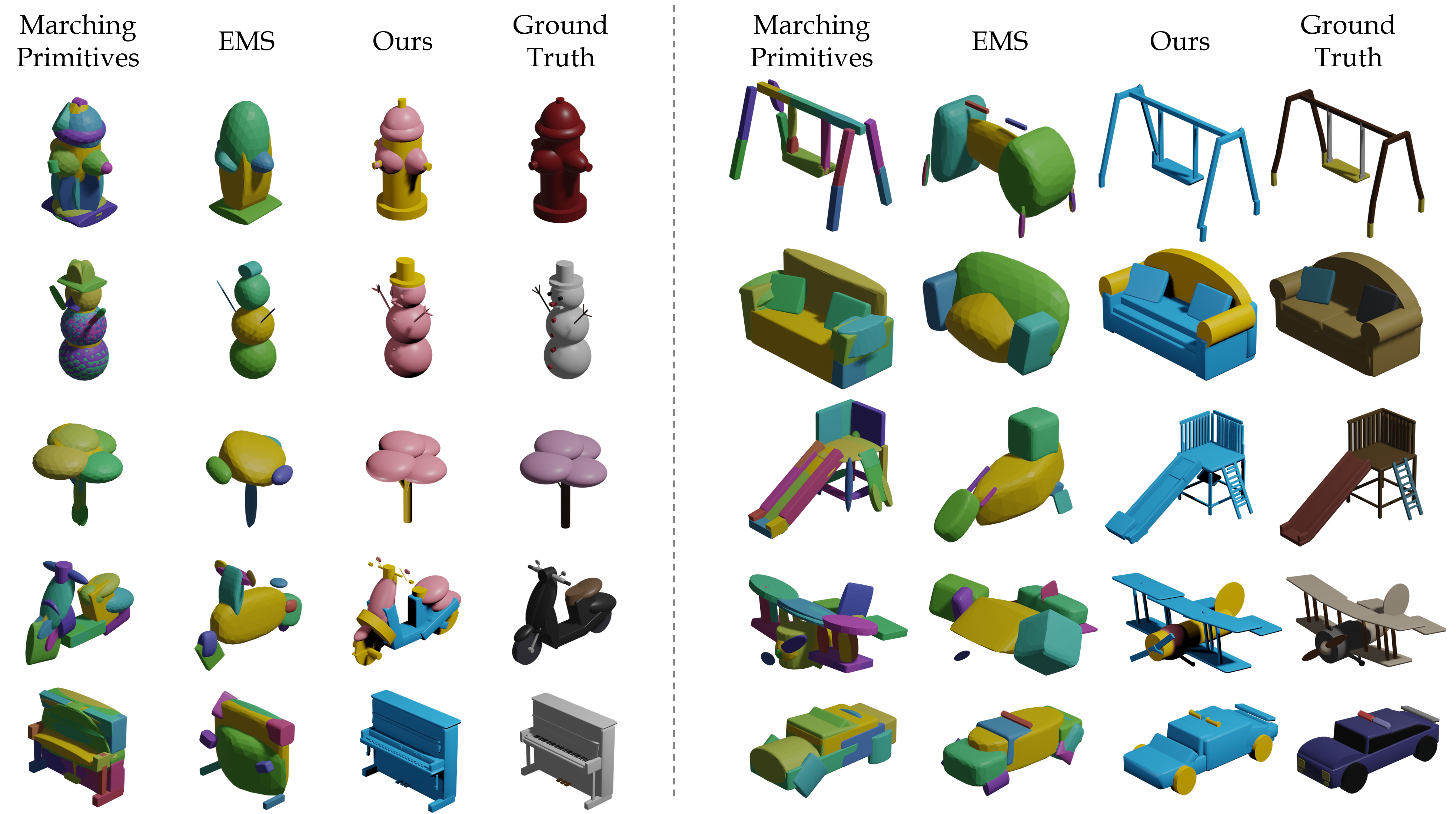}
\caption{More qualitative comparisons with optimization-based methods on the HumanPrim dataset.
}
\label{fig:main2}
\end{figure*}

\begin{figure*}[t]
\centering
\includegraphics[width=0.93\linewidth]{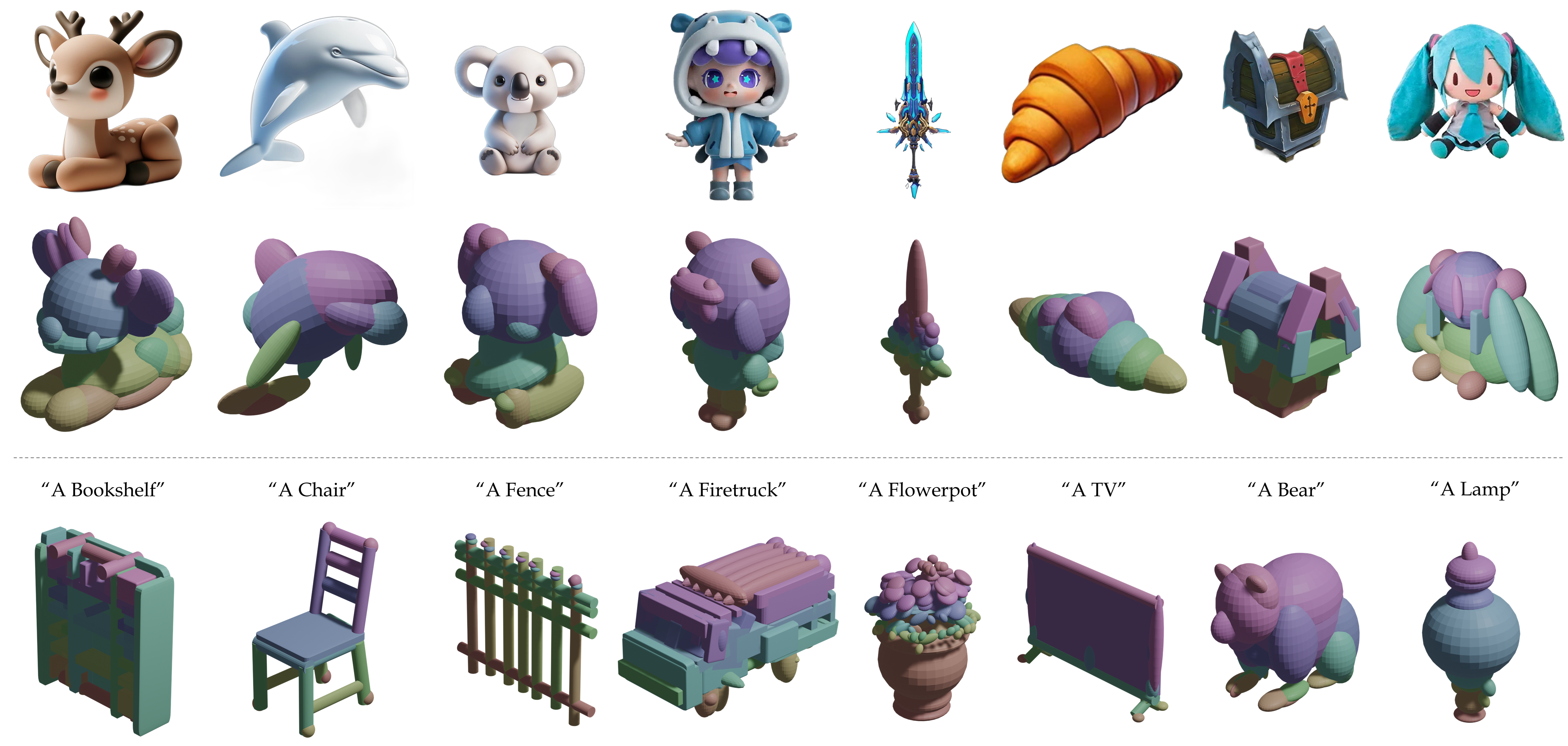}
\caption{
More visualizations of primitive-based 3D content generation on text and image conditions.
}
\label{fig:app2}
\end{figure*}

\subsection{Primitive-based 3D content generation}
Our framework enables versatile primitive-based 3D content generation through its flexible design, which can interface with various 3D generative models to create customized primitive-based 3D content from diverse user inputs, as demonstrated in Fig.~\ref{fig:app} (TRELLIS~\cite{xiang2024structured} for image-conditioning and SDXL~\cite{podell2023sdxl} + Rembg + TRELLIS for text-conditioning).
This approach offers several key advantages over conventional mesh-based representations.
First, since each primitive is directly represented by a predefined primitive type with associated scale, rotation, and translation parameters, users can easily modify the geometric structure through common graphics interfaces while maintaining high modeling capabilities. This accessibility particularly benefits non-expert users in fine-tuning generated results.
Additionally, our primitive-based representation achieves significant storage efficiency, reducing space requirements by over 95\% compared to traditional mesh representations while preserving geometric fidelity.
These characteristics make our method particularly suitable for applications requiring both user interactivity and resource efficiency in 3D content generation.

\section{Conclusion}
In this work, we presented PrimitiveAnything, a novel framework that reformulates 3D shape abstraction as a sequence generation task. 
Our framework learns directly from human-crafted primitive assemblies, enabling it to capture and reproduce intuitive shape decomposition patterns. 
PrimitiveAnything demonstrates strong generalization capability, successfully generating high-quality primitive assemblies across diverse shape categories, enabling versatile primitive-based 3D content creation.
Moreover, the lightweight and efficient nature of primitive representation shows promise for enabling primitive-based user-generated content (UGC) in games.

\begin{acks}
    This work was partially supported by the Natural Science Foundation of China (62461160309) and NSFC-RGC Joint Research Scheme (N\_HKU705/24).
\end{acks}

\bibliographystyle{ACM-Reference-Format}
\bibliography{bib}

\clearpage
\appendix
\setcounter{page}{1}

\section{More Results}
\xhdr{User Study}
To address the human-centric aspects of our method, we conducted a comprehensive user study involving 30 participants (15 female, 15 male) who evaluated 20 randomly selected shapes from the Objaverse dataset. The evaluation focused on three key criteria: (1) geometric similarity, measuring how well the abstraction preserves the original 3D surface structure; (2) anthropomorphism, assessing alignment with human intuition in shape abstraction; and (3) editability, gauging usefulness for interactive editing tasks. Participants rated each abstraction on a 5-point Likert scale (1: poor, 5: excellent). As shown in Tab.~\ref{tab:userstudy}, our method achieved superior average scores across all three metrics compared to EMS and Marching-Primitives. These results validate that our primitive-based shape abstraction scheme not only maintains geometric fidelity but also produces structures that better align with human perception and facilitate easier manipulation for editing tasks.

\begin{table}[htbp]
\centering
\caption{User study results comparing our method with EMS and Marching-Primitives. Each score represents the average rating (on a 5-point scale) from 30 participants evaluating 20 randomly selected Objaverse shapes across three criteria. Our method consistently outperforms baseline approaches.}
\label{tab:userstudy}
\resizebox{1.0\linewidth}{!}{
\begin{tabular}{lccc}
\toprule
\textbf{Method} & \textbf{Geometric Similarity} & \textbf{Anthropomorphism} & \textbf{Editability} \\
\midrule
EMS & 2.16 & 2.18 & 2.17 \\
MP & 3.55 & 3.09 & 3.23 \\
\textbf{Ours} & \textbf{4.17} & \textbf{4.18} & \textbf{4.22} \\
\bottomrule
\end{tabular}
}
\end{table}

\xhdr{Choices of rotation representation}
We opted for Euler angles as the rotation representation in our framework. This decision was driven by several considerations. SVD-based parameterizations, while mathematically elegant, are unsuitable for our application as they lack direct interpolation capabilities, which is crucial for learning-based frameworks. Quaternions, despite their popularity in computer graphics, present challenges including their non-intuitive physical meaning and the requirement for additional constraints (e.g., $w^{2}+x^{2}+y^{2}+z^{2}=1$) to prevent numerical drift, which can complicate implementation and optimization.

Euler angles provide a more intuitive and interpretable representation with their straightforward Euclidean parameter space. Importantly, given that many practical cases in our dataset are nearly gravity-aligned, Euler angles exhibit minimal variance in their values, making the learning process more stable and efficient. To validate our choice, we conducted an empirical comparison between different rotation representations (Quaternions, Rotation Vector, and Euler Angles) as shown in Tab.~\ref{tab:rot}. The results demonstrate that quaternions yield notably inferior results. Rotation vectors perform reasonably well due to their continuity and partial compatibility with gravity-aligned cases, yet still demonstrate a performance gap compared to Euler angles, which consistently deliver superior decomposition quality across our evaluation metrics.

\begin{table}[htbp]
\centering
\caption{Experiments on different choices of rotation representations.}
\resizebox{1.0\linewidth}{!}{
\begin{tabular}{l|c|c|c|c}
\toprule
Representation & CD~$\downarrow$ & EMD~$\downarrow$ & Hausdorff~$\downarrow$ & Voxel-IoU~$\uparrow$ \\
\midrule
Quaternions & 0.0704 & 0.0684 & 0.280 & 0.383 \\
Rotation Vector & 0.0426 & 0.0494 & 0.163 & 0.477 \\
Euler Angles & \textbf{0.0404} & \textbf{0.0475} & \textbf{0.158} & \textbf{0.484} \\
\bottomrule
\end{tabular}
}
\label{tab:rot}
\end{table}

\begin{figure}[t]
\centering
\includegraphics[width=1.0\linewidth]{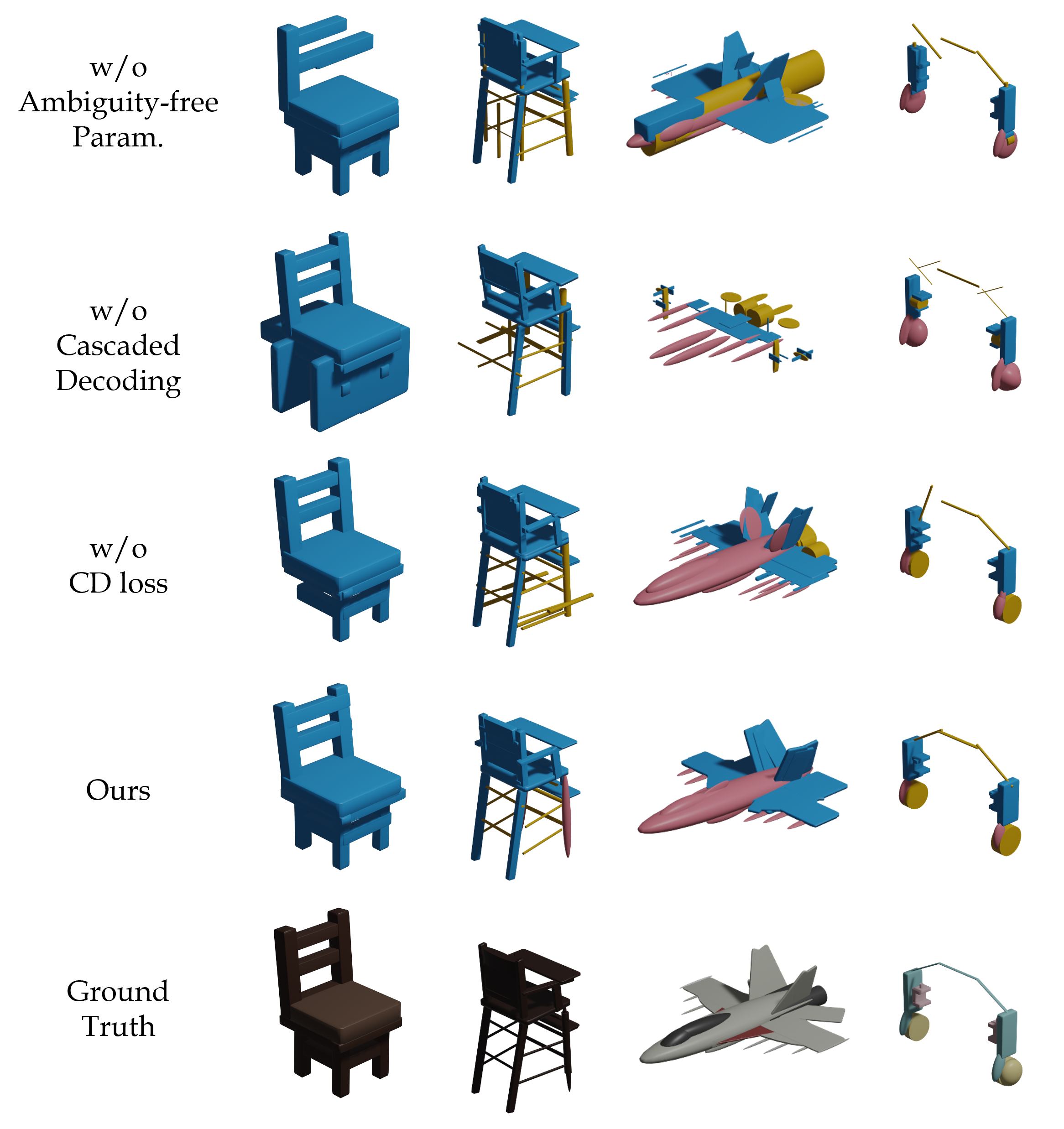}
\caption{Qualitative comparisons of ablation study.}
\label{fig:abl}
\end{figure}

\xhdr{Qualitative results of ablation study}
We provide additional qualitative comparisons in Fig.~\ref{fig:abl} to demonstrate the effectiveness of our ambiguity-free scheme, cascade decoding, and Chamfer distance loss in the ablation study.

\begin{figure*}[htbp]
\centering
\includegraphics[width=1.0\linewidth]{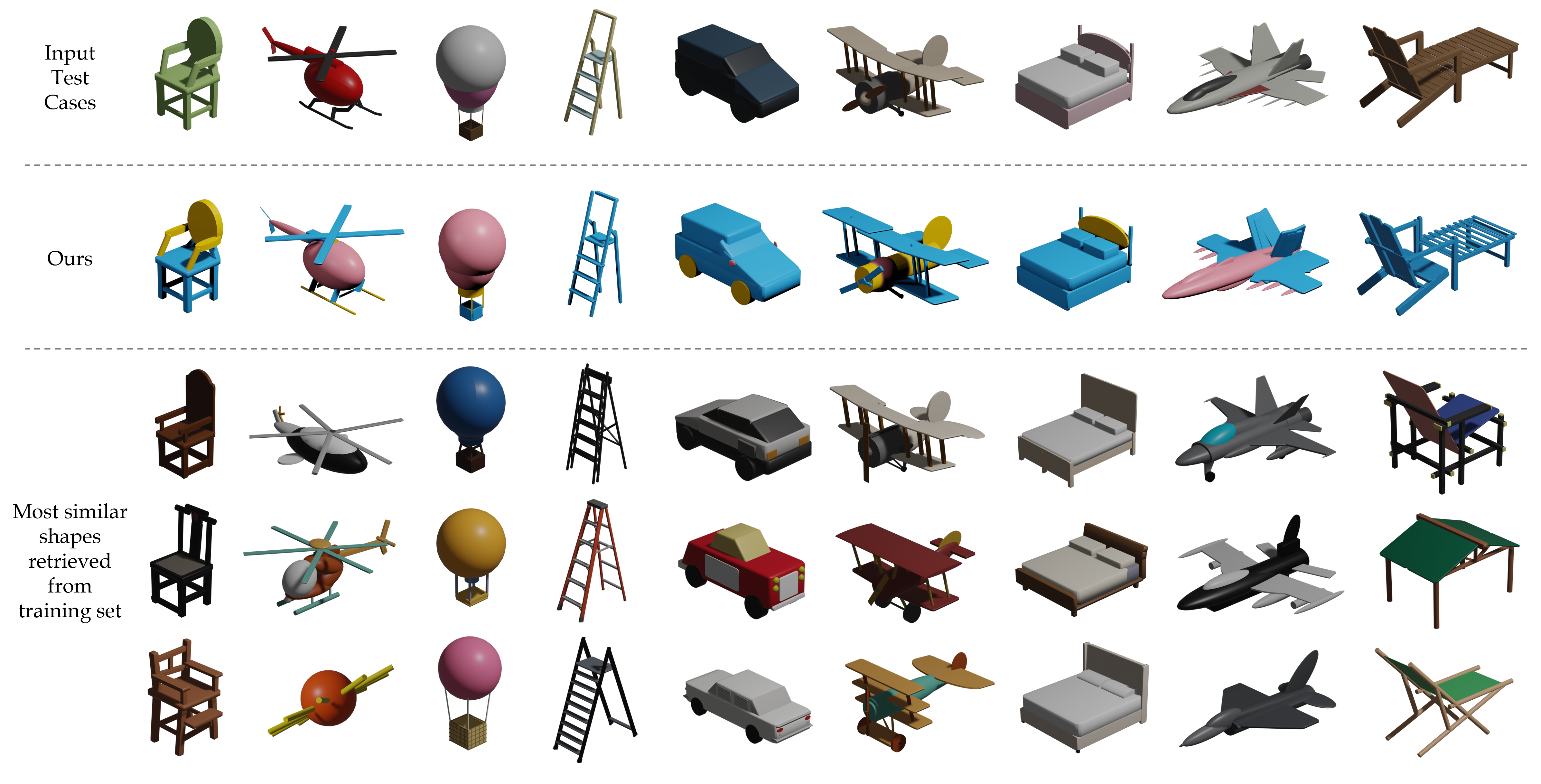}
\caption{Generalization capability of our method. Top row: unseen test shapes; Middle row: our primitive decomposition results; Bottom row: geometrically closest shapes from the training dataset.}
\label{fig:similar}
\end{figure*}

\xhdr{Generalization Analysis}
A key contribution of our method is its ability to perform semantic-aware primitive decomposition on shapes that differ significantly from those in the training data. To empirically validate this claim, we conducted a comparative analysis between test shapes and their geometrically closest counterparts in the training dataset.
We employs the pointbert-vitg14 from the OpenShape~\cite{liu2023openshape} for point cloud feature extraction. We then perform similarity-based object retrieval by measuring the cosine similarity between these extracted feature representations
Fig.~\ref{fig:similar} illustrates this comparison. The top row displays test case shapes that were not seen during training. The middle row shows our method's primitive decomposition results, while the bottom row presents the shapes most similar to those of our training dataset. As evident from the visualization, our method successfully decomposes test shapes into semantically meaningful primitives despite substantial geometric differences from training examples.

\section{More Implementation Details}

\xhdr{Dataset}
To address the need for high-quality shape abstractions with semantic primitives, we carefully constructed a comprehensive dataset through a systematic annotation process. Our annotators were provided with a custom 3D engine featuring an intuitive graphical user interface that enabled precise primitive selection and manipulation (including scale, rotation, and translation operations). Annotators were explicitly instructed to follow two key principles: ensure complete contour coverage of the original shapes and create abstractions that align with human perception. Fig.~\ref{fig:similar} (lower part) showcases representative samples from our dataset.

\begin{figure}[htbp]
\centering
\includegraphics[width=1.0\linewidth]{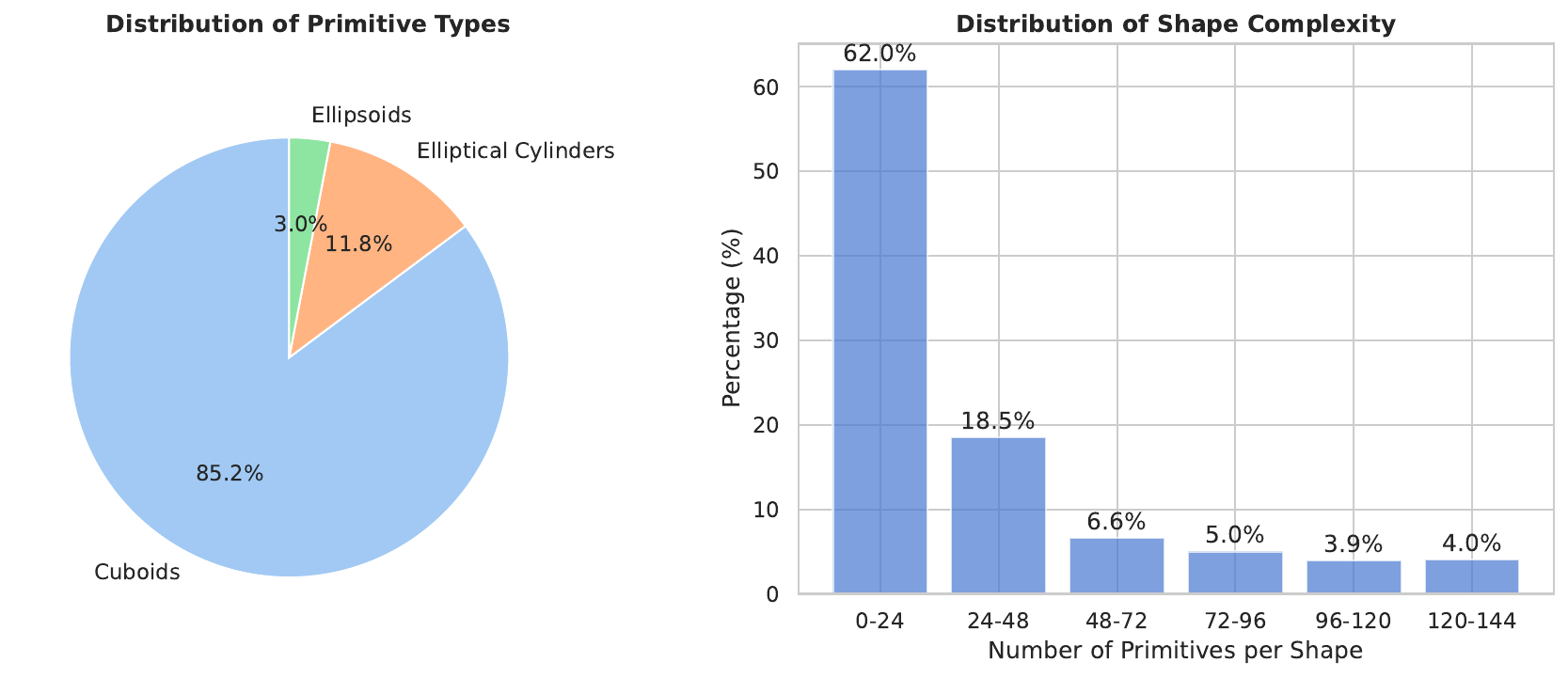}
\caption{Statistical characteristics of our HumanPrim dataset.}
\label{fig:dataset}
\end{figure}

Our HumanPrim dataset exhibits diverse primitive composition characteristics, as shown in Fig.~\ref{fig:dataset}. Analysis reveals that 85.2\% of all primitives are cuboids, 11.8\% are elliptical cylinders, and 3.0\% are ellipsoids. This distribution reflects the predominance of box-like structures in common objects while still incorporating curved surfaces where appropriate.
In terms of complexity, our dataset shows varied primitive counts: data with 0-24, 24-48, 48-72, 72-96, 96-120, and 120-144 primitives account for 62.0\%, 18.5\%, 6.6\%, 5.0\%, 3.9\%, and 4.0\%, respectively. This distribution demonstrates our dataset's balance between simple and complex shape abstractions, supporting robust learning across varying levels of geometric complexity.

\xhdr{Symmetry Order Calculation}
\begin{figure*}[htbp]
\centering
\includegraphics[width=0.85\linewidth]{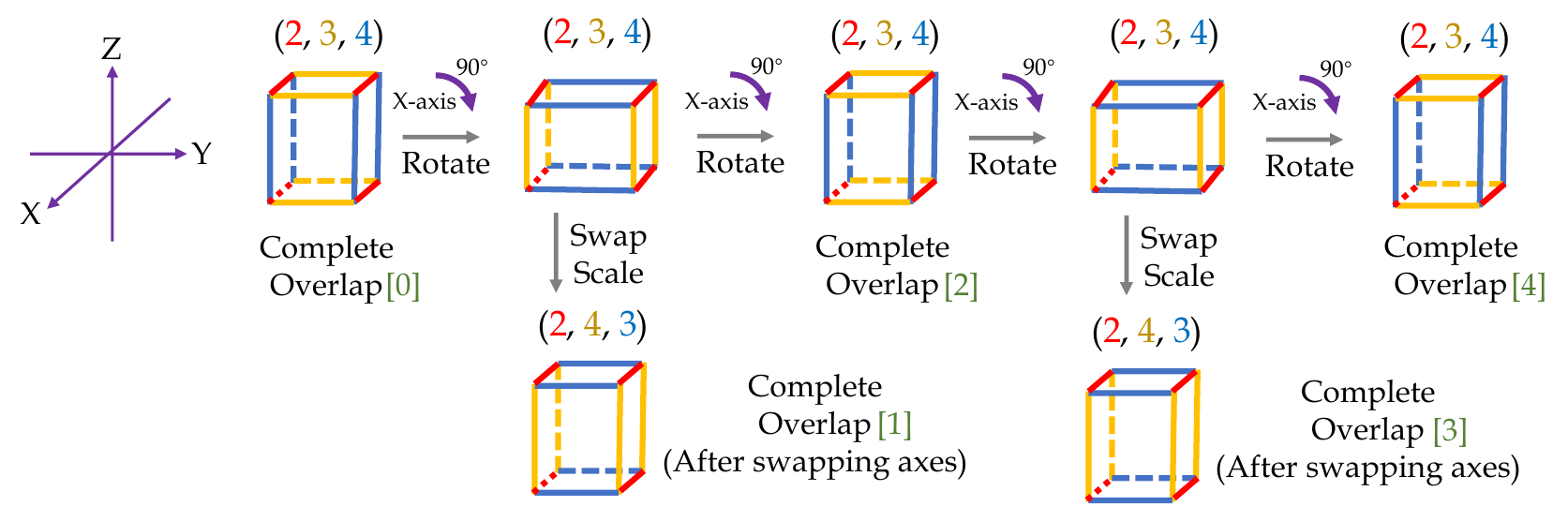}
\caption{Symmetry order calculation for the x-axis of a cuboid. Axis permutation (swapping y and z axes) after 90° and 270° rotations creates configurations equivalent to the original, increasing the symmetry order from 2 to 4.}
\label{fig:order}
\end{figure*}
To determine the total symmetry order of an axis, we account for both rotational symmetry and axis permutations that result in equivalent configurations. This approach captures all possible symmetric transformations of a primitive and is crucial for achieving ambiguity-free parameterization.
Specifically, if swapping two axes after rotation achieves alignment with the original configuration, we include that rotational angle in the symmetry order count.
Fig.~\ref{fig:order} illustrates this calculation process for the x-axis of a cuboid. When considering pure rotational symmetry without axis permutations, only the 180° rotation produces a configuration that perfectly aligns with the original state, yielding a symmetry order of 2. However, our method also recognizes that after 90° and 270° rotations, swapping the y and z axes produces configurations equivalent to the original. By incorporating these axis-permutation-enabled symmetries, the total symmetry order for a cuboid around its x-axis increases to 4.
The explicit inclusion of axis permutations in symmetry calculations parameterizes all possible self-symmetry cases and can apply to all primitives.

\section{More Discussions}

\begin{figure}[htbp]
\centering
\includegraphics[width=1.0\linewidth]{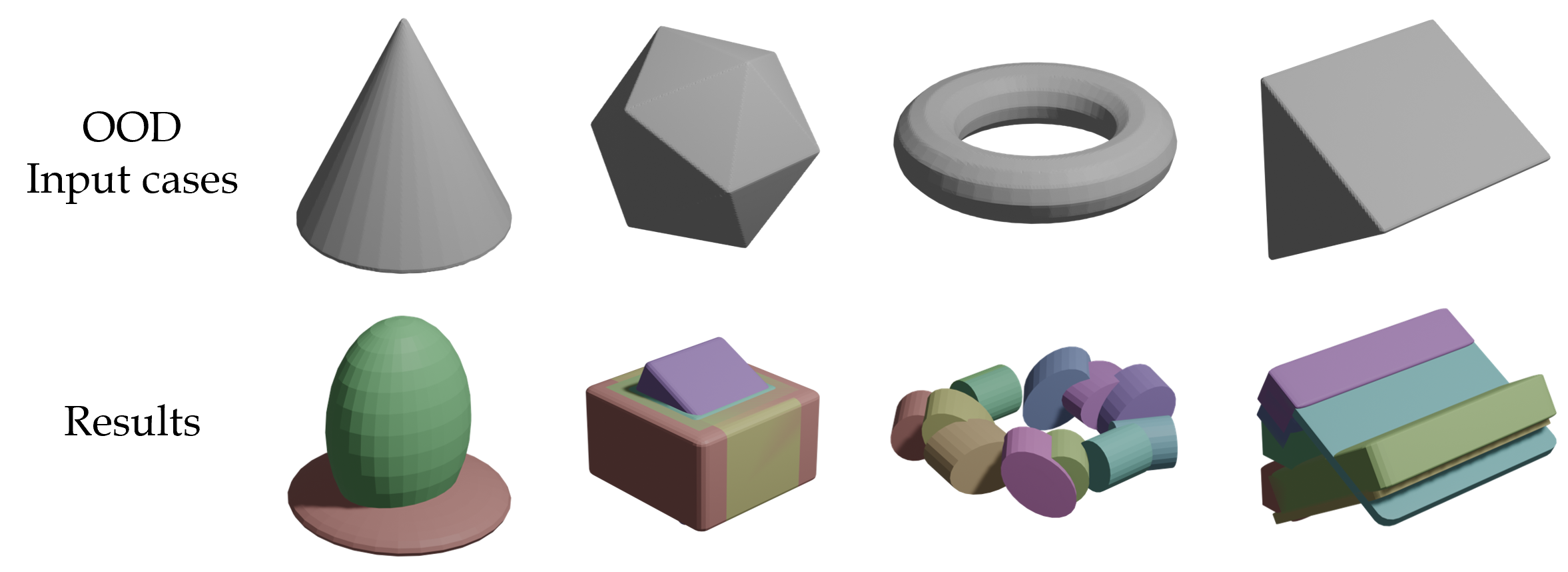}
\caption{Failure cases with out-of-distribution inputs.}
\label{fig:fail}
\end{figure}

\xhdr{Limitations}
Our approach exhibits several limitations despite its effectiveness. Our method struggles with certain out-of-distribution objects, particularly those with topological structures rarely seen in our training data (e.g., ring shapes with holes), as demonstrated in Fig.~\ref{fig:fail}. This challenge stems partly from our current primitive types and could be addressed by expanding our primitive vocabulary and enriching the training dataset with more diverse examples, as our framework is inherently generalizable to such extensions.

We also observe diversity in annotation styles. Some annotators tend to use more primitives than necessary to fully cover the original object, resulting in over-segmentation in certain cases. Regarding design choices of primitive attribute prediction, our discretization scheme focuses learning on larger errors and helps model convergence. However, we acknowledge that this approach introduces trade-offs, potentially causing precision loss and connectivity issues between adjacent primitives.

Symmetry constraints are also excluded to provide greater design freedom, though this sometimes results in asymmetric decompositions of inherently symmetric objects. In addition, our focus remains on geometric shape abstraction rather than appearance modeling. While textures can be handled via back-projection or nearest-neighbour matching with the original 3D object, direct texture generation is not addressed in our pipeline. Both symmetry integration and native texture synthesis represent promising directions for future work that could enhance the practical utility of our method.

\xhdr{Abstraction level of annotations}
The question of appropriate abstraction levels is central to semantic shape decomposition. Annotators were instructed to ensure complete surface coverage while adhering to human-aligned construction principles, which naturally produces varying primitive counts across different 3D objects. This variation reflects inherent complexity rather than enforcing arbitrary consistency.
Unlike optimization-based methods that often fragment semantic parts into multiple pieces with significant overlaps, our approach preserves semantic coherence while maintaining geometric accuracy. This balance stems from our annotation guidelines that prioritize human interpretability alongside geometric fidelity.

Different applications, however, may require different abstraction levels. While detailed decompositions might benefit precise editing tasks, coarser representations could better serve classification or retrieval applications. Though explicitly instructing annotators to provide multiple abstraction levels is challenging, future work could explore inferring these levels based on primitive counts, potentially enabling adaptive abstractions tailored to downstream tasks.

\xhdr{Difference with other abstraction paradigms}
Shape abstraction has evolved along several distinct approaches in the literature, each offering unique perspectives on how to represent 3D objects efficiently and meaningfully. 
Our work contributes to this field through primitive-based shape abstraction, while other paradigms exist, such as hierarchical representations, skeletal abstractions, and surface simplification techniques.

Hierarchical representations like GRASS~\cite{li2017grass} organize shapes into structured trees capturing part relationships and symmetries. While these approaches excel at representing organization, our method offers more direct geometric interpretability through explicit primitive decomposition.
Medial Skeletal Diagram~\cite{guo2024medial} similarly seeks sparse representations, but replaces discrete skeletal elements with continuous primitives, whereas we maintain a clearer separation between structural abstraction (through primitive arrangement) and geometric representation.
Compared to mesh simplification techniques~\cite{QEM} that preserve surface details through vertex/edge removal, our primitive-based abstraction operates at a higher semantic level. We don't just simplify geometry - we reconstruct shapes using fundamental building blocks that naturally align with how humans perceive object structure. This also differs from optimisation-based fitting methods that may over-segment shapes.

\end{document}